\documentstyle[12pt]{article}
        \def\be{\begin{equation}}
        \def\ee{\end{equation}}

\begin{document}
\begin{titlepage}
\vspace*{5mm}

\begin{center} { \bf \large Casimir effect in background of static domain wall} \\

\vskip 1cm

\centerline
\bf M.R. Setare$^{a,}$ \footnote{e-mail: mreza@physics.sharif.ac.ir}
\ \ and \ \
A. Saharian$^{b,}$ \footnote{ e-mail: saharyan@www.physdep.r.am} 

\vskip 1cm

{\it a) Department of Physics, Sharif University of Technology, \\
P.O.Box 11365-9161, Tehran, Iran }\\

\bigskip

{\it b) Department of Physics, Yerevan State University,\\
1 Alex Manoogian St., 375049 Yerevan, Armenia}\\

\end{center}

In this paper we investigate the vacuum expectation values of energy- 
momentum tensor for conformally coupled scalar field in the standard 
parallel plate geometry with Dirichlet boundary conditions and on 
background of planar domain wall case.
First we calculate the vacuum expectation values of energy-momentum
tensor by using the mode sums, then we show that corresponding properties
can be obtained by using the conformal properties of the problem.
The vacuum expectation values of energy-momentum tensor contains two 
terms which come from the boundary conditions and the the gravitational 
background. In the Minkovskian limit our results agree with those obtaind in 
[3].
\end{titlepage}
\newpage

{\bf 1. Introduction}

\bigskip

The Casimir effect is one of the most interesting
manifestations of nontrivial
properties of the vacuum state in quantum field theory
\cite{Plunien}, \cite{Mostepanenko}. Since it's first
prediction by Casimir in 1948 \cite{Casimir48}
this effect is investigated for various
cases of boundary geometries and various types of fields.
The Casimir effect can
be viewed as a polarization of vacuum by boundary conditions. Another type
of vacuum polarization arises in the case of external gravitational field.
In this paper we shall consider a simple example when these two types of
sources for vacuum polarization are present. We investigate
the vacuum expectation
values of the energy-momentum tensor for conformally coupled scalar field
in the standard parallel plate geometry with Dirichlet boundary conditions
and on background of planar static domain wall case.
It has been shown in \cite{Vilenkin1} and \cite{Vilenkin2}, that
the gravitational field of the vacuum domain wall with a source of
the form
\be
T_\mu^\nu=\sigma \delta (x) diag(1,\, 0,\, 1,\, 1) \label{wallemt}
\ee
does not correspond to any exact static solution of Einstein
equations (on domain wall solutions of Einstein-scalar-field
equations see \cite{Widrow}). However the static solutions
can be constructed in presence of an additional background
energy-momentum tensor. Such a type solution has been fond
in \cite{Mansouri}. First we calculate
the vacuum expectation values of
energy-momentum tensor by using the mode sums.
We obtain the result as a direct
sum of two terms: boundary term and term which presents
the vacuum polarization
in the domain wall geometry in the case of absence of boundaries.
It is shown
that boundary part of total energy between the plates and corresponding
pressures on plates are related by standard thermodynamical relation.
Then we show that corresponding properties can be obtained by using the
conformal properties of the problem.

\bigskip

{\bf 2. Vacuum expectation values of energy-momentum tensor}

\bigskip

In this paper we shall consider the conformally coupled real scalar field
$ \phi $, which satisfies
\begin{equation}
( \Box +  \frac {1} {6} R ) \phi = 0 , \qquad 
  \Box =  \frac {1} {\sqrt{ -g }} \partial_{\mu}
( \sqrt{ -g } g^{ \mu \nu } \partial_{\nu} ),  \label{motioneq}
\end{equation}
and propagates on background of gravitational field generated by static
domain wall solution from \cite{Mansouri}.
The corresponding metric has the form
\begin{equation}
ds^2=A^{-2 \alpha} ( dt^2 - dy^2 - dz^2 ) - A^ {-2(\alpha + \gamma + 1)} dx^2 ,      \label{metric}
\qquad A=A(x) \equiv 1+K |x|,  \label{metric}
\end{equation}
where $ \alpha > 0, K>0 $.
The equation (\ref{metric}) describes a planar
domain wall with energy-momentum tensor
$ T_{\mu}^{\nu} =\delta(x) diag(1,0,1,1) $ in the background field with
Ricci tensor
\be
       R_{\mu}^{\nu}=\alpha(\gamma-2\alpha)K^2A^{2(\alpha+\gamma)}diag
       (1,\frac{3\gamma}{\gamma-2\alpha},1,1)        \label{req}
\ee
Note that for the energy density of the background to be positive we must
have $ \gamma< \alpha /2 $.
In what follows as a boundary configuration we shall consider two plates
parallel to each other and to domain wall, with $x$ coordinates equal
to $x_1$ and $x_2$ (to be definit we shall consider right half space of
domain wall geometry $ x_1 , x_2 > 0 $).
For the points on plate the scalar field obeys
Dirichlet boundary condition
\begin{equation}
\phi (x=x_1)=\phi (x=x_2)=0    \label{boundcond}
\end{equation}
The quantization of field (\ref{motioneq})
on background of equation (\ref{metric}) is standard.
Let $\phi _{\alpha }^{(\pm )}(x) $  be complete
set of orthonormalized positive
and negative frequency solutions to the field
equation (\ref{motioneq}), obeying
boundary conditions (\ref{boundcond}). The canonical
quantization can be done by expanding
the general solution of (\ref{motioneq}) in
terms of $\phi _{\alpha }^{(\pm )} $,
\begin{equation}
\phi = \sum_{\alpha } ( \phi _{\alpha }^{+} a_{\alpha } +
\phi _{\alpha } a_{\alpha } ^{(+)}) \label{expansion}
\end{equation}
and declearing the coeficients $a_\alpha $ ,$a_\alpha ^{+} $ as operators
satisfying standard commutation relation
for bosonic fields. The vacuum state
$ |0>$ is defiend as $a_\alpha |0>=0 $. This state is different from the
vacuum state for domain wall geometry
without bondaries, $|\bar 0> $. To investigate
effects due to the presence of boundaries we shall consider vacuum
expectation values of energy-momentum
tensor operator, $<0|T_{\mu \nu }|0> $.
By substituting the expansion (\ref{expansion})
and using the definition of vacuum
state it can be easly seen that \cite{Birrel}
\begin{equation}
<0| T_{\mu \nu } |0>= \sum_{\alpha } T_{\mu \nu } \{
 \phi _{\alpha }^{(+)} ,  \phi _{\alpha }^{(-)}\} \label{vev1}
\end{equation}
Here on the rhs the bilinear form
$ T_{\mu \nu } \{ \phi ,\psi \} $ is determined
by the classical energy-momentum tensor for
conformally coupled scalar field
(see for exampel \cite{Birrel}). To calculate the
vacuum expectation values by (\ref{vev1})
we need the explicit form of eigenfunctions $ \phi _{\alpha }^{(\pm )} $.
For this case the metric and boundary
conditions are static and translation
invariant in directions parallel to the domain wall.
It follows from here
that  the corresponding part has standard plane wave structure:
\be
\phi ^{\pm }_\alpha =\varphi (x)\exp\left [\pm i(k_yy+
k_zz-\omega t)\right ] \label{eigfunc0}
\ee

The equation for $ \varphi (x) $ is obtained from the
field equation (\ref{motioneq}) and for
domain wall metric (\ref{metric}) has the form
\begin{equation}
\varphi ^{''}(x) + ( \gamma +1-2 \alpha ) K sgn(x) A^{-1} \varphi ^{'} +
[ \alpha (\alpha -\gamma )K^{2} A^{-2} +
k_{x}^{2} A^{-2(\gamma +1)}] \varphi =0 \label{eqforx}
\end{equation}
with $ {k_x}^2 =\omega ^2-{k_t}^2 $, $ {k_t}^2={k_y}^2+{k_z}^2 $.
We shall consider the region between plates. The solution
to equation (\ref{eqforx})
in this region obeying boundary conditions (\ref{boundcond}) is
\begin{equation}
\varphi (x)=const A^{\alpha } \sin(k_{x} v), \qquad
k_x=\frac {n \pi }{a}, \qquad n=1,2,... \label{xdep}
\end{equation}
with the relation
\begin{equation}
v(x)=\frac {[ ( 1+ K|x_1|)^{-\gamma } - (1+K |x|)^{-\gamma }]}{K \gamma },  
\qquad a=v(x_2) \label{veq}
\end{equation}
By using this relations and normalizing the eigenfunctions by standard
way one obtains
\begin{equation}
\phi ^{(\pm )}_{\beta }(t, { \vec{r} })= \frac {A^ {\alpha }}
{2 \pi \sqrt{ \omega a}}\sin(k_xv) e^{ \pm i (k_y y + k_z z -\omega t)},
\quad \beta =(n,k_y,k_z), \quad
\omega ^2=\left ( \frac { n \pi }{a} \right )^2+k_{t}^{2} \label{eigfunc2}
\end{equation}
Before to start specific calculation  with
formula (\ref{vev1}) it is convenient to
present energy-momentum tensor for conformally coupeld scalar field in the
form
\begin{equation}
T_{ik}= \partial_i \phi \partial_k \phi - { \frac {1} {6} }
(\nabla _i \nabla _k +
{ \frac {1} {2} } g_{ik} \Box +R_{ik}) \phi ^2 \label{emteq}
\end{equation}
By using the equation (\ref{vev1}) with this form
of energy-momentum tensor and with
eigenmodes (\ref{eigfunc2}) we receive
\begin{equation}
<0|T_{ik}|0>=<0| \partial _{i} \phi \partial _{k} \phi |0>
-{ \frac {1}{6} }
(\nabla _i \nabla _k+ {\frac {1} {2}} g_{ik} \Box +R_{ik})<0|\phi ^2|0>.
\label{emtvev}
\end{equation}
It is convenient first to calculate the quantity \\
$$ <0|\phi (x)\phi (x')|0> =$$
\begin{equation}
\frac{A^\alpha (x) A^\alpha (x')} {4 \pi ^{2}a} \int d^2k_{t}   
\sum_{n=1}^{\infty }
\frac{1}{\omega } \sin(k_xv(x))\sin(k_xv(x')) \\
e^{i[k_y(y-y')+k_z(z-z')-\omega (t-t')]} \label{hadfunc}
\end{equation}
Using Abel-Plana summation formula
\begin{equation}
\sum_{n=1}^{\infty } f(n)= 
\int_{0}^{\infty } f(x)dx-  \frac{1}{2} f(0)+i
\int_{0}^{\infty } \frac{f(ix)-f(-ix)}{e^{2 \pi x}-1} dx
\label{Abel}
\end{equation}
to sum over $n$, and after calculating arising integrals one obtains \\
$$ <0| \phi(x) \phi(x')|0> =
{{ A^{\alpha}(x) A^{\alpha}(x')} \over {8 a^2} }
{\frac {\sinh(u)} {u} }$$
\be
\{ [ \cosh(u) - \cos( \frac{\pi}{a}(v(x)-v(x')) ) ]^{-1} \\
-  [ \cosh(u) - \cos( \frac{\pi}{a}(v(x)+v(x')) ) ]^{-1} \}
\label{hadfunc1}
\ee

where $ u=\frac{\pi}{a}[ (y-y')^2+(z-z')^2-(t-t')^2]^{1/2}$.
The vacuum expectation value of energy-momentum tensor
may be found now by using the relation
\begin{equation}
<0|T_{ik}(x)|0>=lim_{x \to x'} \hat T_{ik} <0| \phi(x) \phi(x') |0>
\label{coinlim}
\end{equation}
where the form of the second order operator $ \hat T_{ik} $
is obvious from (\ref{emtvev}):
\begin{equation}
\hat T_{ik}={ \partial_{i} } { \partial_{k}^{'}} -
       { \frac{1}{6} } ( \nabla_{i} \nabla_{k} + { \frac{1}{2} }
        g_{ik} \Box + R_{ik}) \label{opt}
\end{equation}
and $ \partial_{k}^{'}= \partial / \partial x_{k}^{'} $.
The vacuum expectation value of course is infinite.
This divergencies come
from the first term in figure brackets of (\ref{hadfunc1}).
In this paper we are
mainly interested in quantum effects due to the existence of boundaries.
Let $|\bar 0>$ be the vacuum state for conformally
coupled scalar field in the case of absence of boundaries.
Let us consider the difference

$$ < T_{ik}^{(b)}(x) > = <0|T_{ik}(x)|0>-< \bar{0}| T_{ik}(x)| \bar{0} >= $$
\be
lim_{x \to x'} \hat{T}_{ik}[ <0| \phi(x) \phi(x') |0>
-< \bar{0} | \phi(x) \phi(x')| \bar{0}>]
\label{dif}
\ee
This quantity describes the boundary contribution to the polarization of
the vacuum and is finit. To see this note that the expression for
$ <\bar 0|\phi(x)\phi(x')|\bar 0> $ may be obtained
from (\ref{hadfunc1}) by taking
$ a \to \infty $  and has the following form
\begin{equation}
   <\bar 0|\phi(x)\phi(x')|\bar 0>=-{\frac {A^\alpha(x)
   A^\alpha(x')}{4\pi^2}}\left \{
   [v(x)-v(x')]^2+(y-y')^2+(z-z')^2-(t-t')^2 \right \}
   \label{xhat}
\end{equation}
As it can be easily seen the divergences in (\ref{hadfunc1})
and (\ref{xhat}) cancel in calculating
boundary contribution (\ref{dif}). Substitving (\ref{hadfunc1})
and (\ref{xhat}) into  (\ref{dif}) after some
calculations and arising the second index one
obtains for the region between plates
\begin{equation}
   < T_{i}^{(b)k} > = - \frac {\pi^2 A^{4 \alpha}}{1440 a^4}
   diag(1,-3,1,1), \qquad
   x_1\leq x\leq x_2 \label{diag}
\end{equation}
and zero outside of this region. Here $a$ is defined
by equation (\ref{veq}).
Note that $a$ is different from the proper distance $a_p$ between
the plates,
\be
a_p=\int_{x_1}^{x_2}{g_{11}dx_1}=\left [ (1+Kx_1)^{-\alpha -\gamma}
-(1+Kx_2)^{-\alpha -\gamma }\right ]/k(\alpha +\gamma )
\label{prop}
\ee
In calculating (\ref{diag}) all terms which
come from derivatives $\partial A/\partial x  $
and $\partial ^2A/\partial x^2  $
and are proportional to $ K $ and $ K^2 $ are cancelled. As we shall see
this is direct consequence of the conformal properties
of metric and field
under consideration. In the limit of no geravitation, $K \to 0 $
from (\ref{diag}) we obtain standard Casimir result
for parallel plate configuration.
In the case of scalar field  and $ a=x_2-x_1 $.
By using this result the vacuum expectation value
for total energy-momentum
tensor can be written as
\begin{equation}
<0|T_{ik}|0> =<T_{ik}^{(b)}>+<T_{ik}^{(g)}> \label{totemt}
\end{equation}
where the second summand of right-hand side is the part describing the
polarization of scalar vacuum by domain wall geravitational
field in the case
of absence of boundaries. All divergences are contained in this part.
The corresponding regularization can be done by using the standard methods
of quantum field theory in curved space-time
(see, for example \cite{Birrel}).
Most simply this can be done by using the conformally flatness of the metric
(ref{metric}) (see \cite{Mansouri} and below).
In this cas the anomalous trace determines the total
energy-momentum tensor (see \cite{Birrel}) and the regular part of purely
gravity contribution to (\ref{totemt}) is equal to
\be
< T_{ik}^{(g)} > = reg<\bar 0|T_{ik}^{(g)} |\bar 0> = \nonumber \\
-{ \frac{\alpha+4\gamma}{2880\pi^2}}
\alpha(3\gamma^2-2\alpha^2)A^{4(\alpha+\gamma)}
diag(1,-\frac{3\alpha}{\alpha+4\gamma},1,1) \label{tg}
\ee
now from (\ref{totemt}) and (\ref{tg}) it follows that
regularized total energy-momentum
tensor in the region between plates are given by
\be
< T_{i}^{k} >=<T_{i}^{(b)k} >+< T_{i}^{(g)k}> \label{totemt1}
\ee
where boundary, $ <T_{i}^{(b)k} > $, and geravitational,
 $ < T_{i}^{(g)k} > $, parts
are determined by (\ref{diag}) and (\ref{tg}) respectivly.
In the regions $ x< x_1 $ and
$ x>x_2 $ the boundary part is zero and only gravitational polarization
part remains. The forces acting on plates are determined by boundary part
only. The effective pressure created by
gravitational part in (\ref{totemt1}) is equal to
\be
p_{g1}=-<T_1^{(g)1}>=-\frac{\alpha ^2 K^4(3\gamma ^2-2\alpha ^2)}
{960\pi ^2a^4}A^{4(\alpha +\gamma )}(x) \label{presg}
\ee
and is the same from the both sides of the plates, and hence leads
to the zero effective force. Vacuum boundary part pressures acting
on plates are
\be
p_{b1}^{(1,2)}=p_{b1}(x=x_{1,2})=-<T_1^{(b)1}(x=x_{1,2})>=
-\frac{\pi ^2A^{4\alpha }(x_{1,2})}{480a^4} \label{presb}
\ee
and have attractive nature. The boundary part of the total energy
between the plates can be found by standard way:
\be
E_b=\int_{x_1}^{x_2}\int \int dxdydz\sqrt{-g}<T_0^{(b)0}>=
-\frac{\pi^2}{1440a^3}\int \int dydz \label{toten}
\ee
where we have used the definition of $a$ in accordance with (\ref{veq}).
It can be easily seen that total energy
(\ref{toten}) and pressures (\ref{presb})
are connected by standard thermodynamical relation
\be
p_{b1}(x_1)=-\frac{dE}{dV_1}|_{x_2=const}=-\frac{dE}{dx_1dydz}
A^{4\alpha +\gamma +1}(x_1)|_{x_2=const},\quad
dV_1 =A^{-4\alpha -\gamma -1}(x_1)dx_1dydz \label{thermo}
\ee
and similar relation for $p_{b1}(x_2)$.

Only in the case of conformally invariant fields the eigenmodes have simple
form (\ref{eigfunc2}) in the domain wall geravitational
field (\ref{metric}). This is a direct
consequnce of the conformally equivalence of
the metric (\ref{metric}) to the Minkowskian
one. Indeed by the coordinate transformation
\begin{equation}
x=f(X), \qquad (1+K|f(X)|)^{-\gamma}=1-K\gamma|X| \label{trans}
\end{equation}
It can be seen that the metric (\ref{metric}) takes a
manifastly conformally flat form \cite{Mansouri}.
\begin{equation}
 ds^2=(1-K\gamma|X|)^{\frac{2\alpha}{\gamma}}(dt^2-dX^2-dy^2-dz^2)
 \label{confinv}
\end{equation}
Let $ < T_{ik}^{(M)}> $ be the regularized standard energy-momentum tensor for
a conformally coupled scalar field in the case
of parallel plate counfiguration
in flat space-time with metric $ \eta{ik} $
\begin{equation}
<T_{i}^{(M)k} >=-\frac{\pi^2} {1440 a^4} diag(1,-3,1,1)
\label{minkemt}
\end{equation}
where $ a=x_2-x_1 $ and $ x_i $ are coordinates for the
plates. Using the standard
relation between the energy-momentum tensor
for conformally coupled situation
\begin{equation}
< T_{i}^{k} [ \tilde{g}_{lm} ] > = (
\frac{\eta}{\tilde{g}} )^{ \frac{1}{2}}
< T_{i}^{k(M)} [ \eta_{lm}] > - { \frac{1}{2880 \pi^2} }
\left [ {\frac{1}{6}} ^{(1)}\tilde{H}_{i}^{k}-
^{(3)}\tilde{H}_{i}^{k}\right ] \label{translaw}
\end{equation}
(the standard notations $^{(1,3)}H_i^k$ for some
combinations of curvature tensor components
see \cite{Birrel}), where tilde notes the quantities
in the coordinate system $(t,\, X,\, y,\, z,)$ with metric
(\ref{confinv}). By making the transformation (see (\ref{trans}))
to the initial coordinate system $(t,\, x,\, y,\, z)$ from
(\ref{translaw}) we receive the result (\ref{totemt1}), where
$a$ is expressed via the coordinates $x_1,\, x_2$ of plates
in system (\ref{metric}) by relation (\ref{veq}).\\

{\bf 3. Concluding remarks}

In this paper we calculate the Casimir energy for conformally invariant
scalar field in the standard parallel plate, on background of planar
static domain wall. The boundary conditions over scalar field on the plate
are Dirichlet boundary conditions.
For calculating the vacuum expectation values of the energy-momentum tensor
we use the mode sums method and Abel-Plana summation formula. The result
contains two terms, one comes from the boundary conditions and the other one
from the effect of gravitation over the vacuum of scalar filed. The quantity
which reflects the effect of bondary conditions is finit. If we write this
term in flat space-time limit we obtain the standard result of casimir effect
for parallel plates.
All divergences are in the part which describs the polarization of scalar
vacuum by domain wall background in case of absence of boundaries.
The effective pressure created by gravitational part is the same for both
sides of the plates and hence leads to the zero effective force, but the
vacuum boundary part pressures acting on plates is attractive. When the
total energy between the plates are found, one can readily see that the
total energy and pressures are connected through the standard thermodynamical
relations.
These results can easily be obtained using the conformal properties of the
metric and the scalar field. \\

{\bf Acknowledgment } \\
M. R. Setare would like to thank R. Mansouri for his valuable hints
and comments. He also thanks F.H. Jafarpour for his help in
preparing the Latex file.


\begin{thebibliography}{99}

\bibitem{Plunien} G. Plunien, B. Mueller, W. Greiner, Phys.Rep.
{\bf 134}, 87 (1986).

\bibitem{Mostepanenko} V. M. Mostepanenko, N. N. Trunov,
Sov.Phys.Usp.31 (11) November 1988
The Casimir effect and it's applications.

\bibitem{Casimir48} H. B. G. Casimir, Proc. K. Ned. Akad. Wet.
{\bf 51},793 (1948).

\bibitem{Vilenkin1} A. Vilenkin, Phys.Rev. {\bf D23}, 852 (1981).

\bibitem{Vilenkin2} A. Vilenkin, Phys.Lett. {\bf 133B}, 177 (1983).

\bibitem{Widrow} L. M. Widrow, Phys.Rev. {\bf D39}, 3571 (1989).

\bibitem{Mansouri} R. Mansouri, Gravitational field of static
domain wall, Preprint UWThPh-1989-12.

\bibitem{Birrel} N. D. Birrel, P. C. W. Davies, Quantum fields
in curved space. Cambridge University Press, Cambridge 1986.

\end{thebibliography}
\end{document}